\documentclass[11pt]{article}
\usepackage[psamsfonts]{amssymb}
\usepackage{amsmath}

\newcommand{\ed}{\end{document}}
\newcommand{\beq}{\begin{eqnarray}}
\newcommand{\eeq}{\end{eqnarray}}
\newcommand{\beqz}{\begin{eqnarray*}}
\newcommand{\eeqz}{\end{eqnarray*}}
\newcommand{\nn}{\nonumber }

\newcommand{\bP}{\mathbf P}
\newcommand{\bE}{\mathbf E}
\usepackage{color}

\begin{document}

\title{Quantum Decision, Quantum Logic,\\and Fuzzy sets}
\author{Grigorii Melnichenko\\
Vilnius Pedagogical University, Studentu St. 39\\
LT-08106 Vilnius, Lithuania\\
e-mail: gmelnicenko@gmail.com, grimel@vpu.lt}

\date{}

\maketitle

\begin{abstract}
In the paper, we show that quantum logic of linear subspaces can be used for recognition of random signals by a Bayesian energy discriminant classifier. The energy distribution on linear subspaces is described by the correlation matrix of the probability distribution. We show that the correlation matrix corresponds to von Neumann density matrix in  quantum theory. We suggest the interpretation of quantum logic as a fuzzy logic of fuzzy sets. The use of quantum logic for recognition is based on the fact that the probability distribution of each class lies approximately in a lower-dimensional subspace of feature space. We offer the interpretation of discriminant functions as membership functions of fuzzy sets. Also we offer the quality functional for optimal choice  of discriminant functions for recognition from some class of discriminant functions.

\medskip
{\it Key~words:} recognition, quantum logic, discriminant function,
fuzzy set, von Neumann density matrix, membership functions, subspace
classifier, quality functional, quantum decision.
\end{abstract}

\section{Introduction}

A Bayesian probabilistic discriminant classifier is based on a classical
probability theory using algebra of subsets.
The decision rule of the probabilistic classifier maximizes
the probability of ``correct'' recognition. A Bayesian energy discriminant
classifier was briefly presented in [12]. The algebra of linear
subspaces (quantum logic) is used instead of algebra of subsets. The
decision rule of energy classifier maximizes the energy of ``correct''
recognition. The recognition of two classes is considered in detail. The
use of quantum logic for recognition of signals is considered in [10].

   The use of linear subspaces as class models is based on the assumption
that the distribution of each class lies approximately in a
lower-dimensional subspace of  feature space. These spaces can be found by
principal components analysis carried out individually on each class. An
input vector from the unknown class is classified according to the
greatest projection to the subspaces, each of which represents one
class.

   The subspace classifier was suggested  by Watanabe (method CLAFIC [3], [4]).
This method,  however, has drawbacks: a priori probabilities of classes are
not used; subspaces of classes can  overlap. T. Kohonen has
offered the Learning Subspace Method (LSM) [2], [3]. During the training LSM
decreases the number of vectors that are included in subspaces of
different classes. The recognition of handwritten signs by the subspace
classifier is considered in [4]. The subspace classifier is applied to
phonemes recognition in [5] and to speaker recognition in [6].

   Y.C.~Eldar and A.V.~Oppenheim~[7] draw a parallel between quantum measurements
 and algorithms in signal processing. They propose to exploit the rich
mathematical structure of quantum theory in signal processing without
realization of quantum processes. We suggest to consider energy
processes instead of quantum processes because nature spends some energy
to create any signal.

\section{Quantum logic as an example of fuzzy logic}

    Let $H$ be a Hilbert space. A fuzzy set $A$ of $H$
 is a set of ordered pairs
 $A = \{ {x, \mu _A ( x )\colon\ x \in H} \}$ where
 $\mu_A (x)$: $H \to\{ {0,\infty} \}$ is the membership function of
 the fuzzy set $A$. Suppose $\mu_A ( x )$ be non necessarily normal:
 $\sup\mu_A ( x ) \ne 1, x \in\mathrm{H}$.
 A set of membership functions is a partially ordered
 set equipped with a partial order relation:
 $\mu_A ( x ) \le\mu_B ( x )$ for all $x \in H$.
 The result of operations
$$
\mu_A ( x ) \wedge\mu_B ( x ) = \inf\big ( {\mu _A ( x ),\mu _B ( x )}\big
),\quad
\mu_A ( x ) \vee\mu_B ( x ) = \sup\big ( {\mu _A ( x ),\mu _B ( x )} \big )
$$
 is defined pointwise and the result is again a
 nonnegative function. Hence, the set of membership functions is a lattice.

    Each closed linear subspace $M \subset H$ corresponds to an elementary
 logical proposition of quantum logic. Each linear subspace $M$ has an
 orthogonal projection $P_M$ onto $M$. So a proposition of quantum
 logic can be associated with the orthogonal projection. The set of all
 orthogonal projections is a lattice equipped with a partial order relation:
 $P \le R$ if $\langle{Px,x} \rangle \le\langle{Rx,x} \rangle$
 for all $x \in H$. Hence every pair of projections $P, R$ has a unique
 supremum (least upper bound) and a unique infimum (greatest lower bound):
$$
P \wedge R = \inf( {P,R} ),\quad  P \vee R = \sup(P,R).
$$
 Operations $P \wedge R$,  $P \vee R$,  and $P^ \bot = I - P$ are
 conjunction, disjunction, and negation of quantum logic, respectively.

    Each projection $P_M$ on the subspace $M$ can be viewed as a filter [10]
 and it passes some energy
 $\mu_M ( x ) = \langle{P_M x,x} \rangle = ||P_M x||^2$ of signal $x$
 (in quantum theory, a projection passes some quantum probability.
 This energy evaluates the value of membership of signal $x$ to subspace $M$.
 So each linear subspace $M \subset H$ can be associated with the fuzzy set:
$$
A_M = \big \{ {x, \mu _M ( x )\colon\ x \in H,\ M \subset H} \big \},\quad
\mbox{where}\ \ \mu_M ( x) = \langle{P_M x,x} \rangle .
$$

     A set of all membership functions $\{ {\mu _M ( x ), M \in H} \}$
 is a lattice equipped with a partial order relation:
 $\langle{P_M x,x}\rangle\le\langle{P_N x,x} \rangle$ for all $x \in H$.
 So operations supremum and infimum of that lattice can be used as
 a fuzzy logic conjunction and disjunction of fuzzy sets
 $\{ {A_M, M \in H} \}$. A fuzzy logic negation of fuzzy set $A_M$
 with membership function $\mu_M ( x )$ can be defined as a fuzzy set
 $A_{M^\bot}$ using the following membership function:
 $\mu_{M^ \bot}(x )= \langle{ {P_{M^\bot}} x,x} \rangle =
 \langle{ {P_M^\bot} x,x} \rangle =
 \langle{( {I - P_M} )x,x} \rangle$, where a subspace $M^ \bot$ is an orthogonal
 complement of subspace $M$.
 Thus fuzzy sets $\{ {A_M,M \in H} \}$ form a fuzzy logic.

\section{Discriminant functions as membership functions}

If an object of recognition is described as a vector $x = ( {x_1,  x_1,\ldots, x_n} )$, then the vector $x$ is the pattern of the object
in the feature space $H = R^n$.  A membership of object to some class
$S_i$, $i = 1\ldots l$, is an additional feature, which can be
defined as the index $i$ of the class, where $i \in I= \{ 1, 2,\ldots, l\}$.

We use discriminant functions for the classifier of recognition.
Discriminant functions are a set of functions $g_i ( x )$, $i = 1 \ldots l,$
that determine the membership of the object with the pattern $x$ to some
class $S_i$ according to the following decision rule: if the object with
the pattern $x$ satisfies $g_i ( x ) > g_j ( x )$ for all $j \ne i$,
then the object having the pattern $x$ belongs to the class~$S_i$.

    Discriminant functions split the feature space $H$ into disjoint sets:
\[
A_i = \big \{ {x\colon\ g_i (x) > g_j (x), j =
1\ldots l,  \ j \ne i} \big \}.
\]
    Thus, if $x \in A_i$,  then the object having the pattern $x$ belongs to
 the class $S_i$.  However, there are sets
 $\{ {x\colon\ g_i ( x ) = g_j ( x ),\ j \ne i} \}$,  $i = 1\ldots l$,
 whose elements it is impossible to include in some set $A_i$,  $i = 1 \ldots l$.
 Usually these sets are included in the  sets
 $A_i$,  $i = 1\ldots {\kern 1pt} l$.

 Using discriminant
 functions, the classifier determines only a ``likehood'' value about
 the membership of the object with the pattern $x$ to some class $S_i$.
 So discriminant functions $g_i ( x )$, $i = 1 \ldots l$, are
 membership functions. In the following, we assume that discriminant
 functions are negative and non-necessarily normal:
 $\sup g_i ( x ) \ne 1$, $x \in H$,  $i = 1\ldots l$.

\section{Quality functional for a choice of optimal decision rule}

 We shall use a probabilistic model for  recognition. Let
 $( {\Omega,\mathcal{A},\bP} )$ be a probability space where a sample space
 $\Omega$ is a set of recognition objects. It is evident that the set
 of recognition classes $S_1,  S_2,  \ldots , S_l$ are a partition of
 $\Omega $: $S_1 + S_2 + \cdots + S_l = \Omega$ where
 $S_i \cap S_j = \emptyset$ for all $i \ne j$.

    Following Zadeh [1], a fuzzy set $A$ is called
 a fuzzy event if the corresponding membership function
 $\mu_A ( \omega)$: $\Omega\to\{{0,\infty} \}$ is $\mathcal{A} $-measurable.
 The probability of a fuzzy event is defined as
\beq
\bP( A ) = \bE\mu_A = \int\limits_\Omega {\mu _A \mathrm{d} \mathrm{P}}.
\eeq

    Suppose  that an object $\omega$ is described by the vector
 $\xi( \omega) = ({\xi _1( \omega ),\xi _2 ( \omega ), \ldots, \xi _n ( \omega )} )$
 of features where each
 $\xi_i ( \omega)$: $\Omega\to H$, $i = 1\ldots l$, is $\mathcal{A}$-measurable
 random variable. Since an object $\omega$ has the pattern $x$ in the feature space
 $H$, there is a map $\xi( \omega)$: $\Omega\to H$. If $\omega\in S_i$,
 then we can define an integer-valued random variable $\gamma$ such that
 $\gamma( \omega) = i$ for all $\omega\in S_i$, where
 $i \in I\mathrm{ = }\{ {1,2,\ldots, l} \}$.
 The sample space $\Omega$ of the objects usually not accessible to immediate
 observation, therefore it is necessary to deal with the feature space $H$.
 However, $\Omega$ can be identified with $I \times H$.

    We use a Bayesian method which needs a priory probabilities
 $p_i = \bP({S_i} )$, $i = 1\ldots l$, and a conditional distributions
 $\mu_i ( A ) = \bP( {\xi \in A| {S_i} } ), i = 1\ldots l$. Since
 $\bP( {S_i} ) = \bP( {\gamma = i} )$, it follows that
 $p_i , i = 1\ldots  l$, is the probability distribution of the
 random variable $\gamma$.

Let $\mu( {B,A} ) = \bP( {\gamma \in B,  \xi \in A} )$ be a joint
distribution of random variables $\gamma $, $\xi$, where $B = \{
i_1, i_2,  \ldots ,i_m \} \subset I$ and $A \subset\mathrm{H}$.  We have
$\mu_i ( A ) = \bP( {\xi \in A| {S_i} } )$, $i = 1\ldots l$.  Since  $S_i = ( {\gamma = i} )$,  we get
$$
\mu( {\{ i \} ,A} ) =\bP( {\gamma = i, \xi \in A} ) = \bP(\xi \in A |\gamma =
i)\bP(\gamma = i) = \bP(\xi \in A |{S_i})\bP({S_i}) = p_i \mu_i (A).
$$
Let us denote $\mu_1(\{ i\}) = p_i$ and $\mu^1_2( {i,A} ) = \mu_i ( A )$.  We have
\beqz
\mu (B, A) &=&
\bP\bigg ( \sum_{k = 1}^m(\gamma = i_k) \cap (\xi \in A)\bigg )=
\sum_{k = 1}^m \bP(\xi \in A\,|\,{\gamma = i_k})\bP(\gamma = i_k ) \\
&=&
\sum_{k =1}^m \frac{\mu (\{ i_k \}, A )}
{p_{i_k}} p_{i_k} =
\sum_{k = 1}^m  \mu^1_2 ( {i_k, A} )\mu_1(\{ i_k\}) =
\int\limits_B \mu^1_2({i,A})\mu_1(\mathrm{d}\,i).
\eeqz
It follows that $\mu^1_2( {i,A} ) = \mu_i ( A )$ is the transition
probability on $I \times \mathcal{ B }$~[11], where $\mathcal{ B }$ is a
$\sigma$-algebra of Borel subsets of feature space $H = R^n$.

    Discriminant functions $g_i ( x )$, $i = 1\ldots l$, define a
random variable $g_\gamma( \xi) = g( {\gamma, \xi} )$.
 Since $\mu^1_2( {i,A} ) = \mu_i ( A )$ is the transition probability on
  $I \times \mathcal{ B } $~[11], we have

\beq
\bE g( {\gamma,  \xi} ) = \int\limits_I {\mu_1( {\mathrm{d}{\kern
1pt} i} )} \int\limits_H {g( {i,x} ) \mu^1_2 ( {i,\mathrm{d}{\kern 1pt}
x} )} = \sum_{i = 1}^l {p_i \int\limits_H {g_i ( x )\mu _i (
{\mathrm{d}{\kern 1pt} x} )}}.
\eeq

Suppose $H = A_1 + A_2 + \cdots + A{\kern 1pt} _l$,  where $A_i, {\kern
2pt} i = 1\ldots l,$ are disjoint sets. Let $\Phi$ be a class of
discriminant functions which contain only indicator functions:
$$
g_i (x) = 1_{A{\kern 1pt} _i} (x) =
\begin{cases}
1 & \text{if\  $x \in A{\kern 1pt} _i$}, \cr
0 & \text{if\  $x \notin A{\kern 1pt} _i$}. \cr
\end{cases}
$$
It is evident that $g_{\gamma ( \omega )} ( {\xi ( \omega )} ) = g(
{\gamma ( \omega ),\xi ( \omega )} )$ is the indicator function with a
support:
$$
G = \sum_{i = 1}^l {( {\xi \in A_i} ) \cap (
{\gamma = i} )} = \sum_{i = 1}^l {( {\xi \in A_i}
) \cap S_i}.
$$

We can say that the indicator function $1_G = g( {\gamma, \xi} )$ is the
membership function of ``correct'' recognition, where $G$ is a crisp
event  of ``correct'' recognition. By (2), we have
\beq
\bP( G ) = \bE g( {\gamma, \xi} ) = \sum_{i = 1}^l {p_i \int\limits_H
{g( {i,x} )}} \mu_i ( {\mathrm{d}x} ) = \sum_{i = 1}^l {\bP( {\xi
\in A_i | {S_i}} )\bP( {S_i} )}.
\eeq

A Bayesian probabilistic discriminant classifier splits the feature space
$H$ on disjoint sets $H = A_1 + A_2 + \cdots + A{\kern 1pt} _l$ such
that the probability (3) for the crisp event $G$ of ``correct'' recognition
would be maximal.

Let $g_i ( x ), i = 1\ldots l$, be discriminant functions from
some class $\Phi$, where each function $g_i ( x )$: $H \to\{ {0,\infty} \}$
is a Borel-measurable membership function of class $S_i$.  Then the random
variable $g_i ( {\xi ( \omega )} )$, $i = 1\ldots l$  on $\Omega$ is a membership
function such that the value $g_i ( {\xi ( \omega )} )$ is a
membership degree of object $\omega$ to a class $S_i$.  We define a
fuzzy event as follows: $G_i = \{ {\omega,  g_i ( {\gamma ( \omega ),\xi ( \omega
)} )\colon\ \omega \in \Omega} \}$ for all $i = 1\ldots l$.

Let us define the membership function:
$$
\mu_j ( {i,\omega} ) = 1_{S_j} ( \omega) g_i
\big (\xi ( \omega )\big ) = \begin{cases}
g_i\big  (\xi ( \omega )\big  ) & \text{if\ \ $\omega \in S_j$,} \\
0 & \text{if\ \ $\omega \notin S_j$. } \\
\end{cases}
$$
This membership function defines the fuzzy event $S_j G_i = \{ {\omega,
\mu _j ( {i,\omega} )\colon\ \omega \in \Omega} \},$ which is an algebraic product [1] of
events $G_j$ and $S_i$. The value   $\mu_j ( {i,\omega} )$ is
the membership degree of the object $\omega$ to the class $S_i$ if the statement $\omega\in S_j$ is true. There can be two cases. First, if $j = i$, then $\mu_i ( {i,\omega} )$
is the membership degree of the object $\omega$ to the class $S_i$  when the object $\omega$
belongs to its own class $S_i$. We call the value $\mu_i ( {i,\omega}
)$  a ``correct'' degree of membership; we call the fuzzy event
$S_i G_i$   a fuzzy event of ``correct'' recognition. Second,
if $j \ne i$, then $\mu_j ( {i,\omega} )$ is the membership degree of the
object $\omega$ to the class $S_i$
when the object $\omega$ belongs to other class $S_j$.  We call the
value $\mu_j ( {i,\omega} )$, $j \ne i$, an ``error'' degree of
membership; we call the fuzzy event $S_j G_i$,  $j \ne i$,  a fuzzy
event of ``error'' recognition.

Since  $1_{S_i} = 1_{( {\gamma = i} )}$ for all $i = 1\ldots l$,  we can
define a membership function:
$$
g( {\gamma, \xi} )= \sum_{i = 1}^l {1_{( {\gamma
= i} )} g ( {\gamma, \xi} )} = \sum_{i = 1}^l {1_{( {\gamma
= i} )} g_i ( \xi )} = \sum_{i = 1}^l {1_{S_i}
g_i ( \xi )} = \sum_{i = 1}^l {\mu _j (
{i,\omega} )}.
$$
This membership function defines a degree of ``correct'' membership
for all objects $\omega\in\Omega$. We call the random variable $g( {\gamma
,\xi} )$ as a membership function of ``correct'' recognition and the
fuzzy set $G = \{ {\omega,  g( {\gamma ( \omega ),\xi ( \omega )}
)\colon\ \omega \in \Omega} \}$  as a fuzzy event of ``correct'' recognition.

   It is natural to choose discriminant functions $g_i ( x )$, $i = 1\ldots l$
from the class $\Phi$ such that the probability of the fuzzy event $G$ of
``correct'' recognition would be maximal. From (1) and (2), we have that
the probability of the fuzzy event $G$ is defined as
\beq
\bP( G ) = \bE g( {\gamma, \xi} ) = \sum_{i = 1}^l {p_i \int\limits_H
{g( {i,x} )}} \mu_i ( {\mathrm{d}x} ).
\eeq
Also (4) defines a quality functional for choice of discriminant
functions from the class~$\Phi$.

Let us show  another   interpretation of the quality functional (4). We define
$$
1_{( {k = j} )} ( k ) = \begin{cases}
1 & \text{if\ \ $k = j$,}  \\
0 & \text{if\ \ $k \ne j$.}  \\
\end{cases}
$$
Let us denote  $\mu_1(\{ i\}) = p_i$. Since
 $1_{S_i} = 1_{( {\gamma = j} )}$ and $\mu^1_2( {i,A} )=\mu_i ( A )$
 is a transition probability on $I \times \mathcal{ B } $, it follows that [11]
\beqz
\bE\big ( 1_{S_j} g_i ( \xi )\big )&=&
\bE\big (1_{(\gamma = j)} g_i ( \xi )\big ) \\
&=&\mathop{\int\,\int}\limits_{I \times H}
1_{(k = j)}g_i ( x) \mu (\mathrm{d}{\kern 1pt}k,\mathrm{d}{\kern 1pt}
x) =\int\limits _I\mu_1(\mathrm{d}{\kern 1pt}k)1_{(k = j)}\int\limits_H g_i
(x)\mu^1_2 (k,\mathrm{d}{\kern 1pt} x )\\
&=&
\sum_{k = 1}^l \mu_1(\{ k\})  1_{(k = j)} \int\limits_H g_i
(x)\mu^1_2 (k,\mathrm{d}{\kern 1pt} x ) =
p_j \int\limits_H g_i (x)\mu _j (\mathrm{d}{\kern 1pt} x).
\eeqz
Then the probability of the fuzzy event $S_j G_i = \{ {\omega,  1_{S_j} (
\omega ) g_i ( {\xi ( \omega )} )\colon\ \omega \in \Omega} \}$ is defined as
\beq
r_j ( i ) = \bP( {S_j G_i} ) = \bE\big ( {1_{S_j} g_i ( \xi )}\big  ) = p_j
\int\limits_H {g_i ( x )\mu _j ( {\mathrm{d}{\kern 1pt} x} )}.
\eeq
We call the value $r_j ( i )$  a ``correct'' probability of
recognition if $i = j$ and  an ``error'' probability of recognition if
$i \ne j$.  The full sum of all the ``correct'' probability of recognition
is defined as
$$
\sum_{i = 1}^l {r_i ( i )} = \sum_{i = 1}^l
{\bP( {S_i G_i} )} = \sum_{k = 1}^l {\bE\big ( {1_{S_i}
g_i ( \xi )} \big )} = \sum_{i = 1}^l {p_i
\int\limits_H {g_i (x)\mu _i ( {\mathrm{d}x} )}}
= \bE g( {\gamma, \xi} ) = \bP( G ).
$$

    Let us define a conditional expectation of random variable relative to
an event:
$$
\bE\big ( {g_i ( \xi )1_{S_i}} \big ) = \frac{\bE( {g_i ( \xi )1_{S_i}}
)}{\bP( {S_i})}\bP( {S_i} ) = \bE\big ( {g_i ( { \xi} )| {S_i} }\big )\bP(
{S_i} ),\quad \mbox{where}\ \ i =1\ldots l.
$$
Then we get one more interpretation of the quality functional (4):
$$
\bP( G ) = \bE g( {\gamma,  \xi} ) = \bE\bigg ( {\sum_{i = 1}^l {1_{( {\gamma
= i} )} g( {\gamma,  \xi} )}} \bigg ) = \sum_{i = 1}^l {\bE\big ( {1_{S_i} g_i
( \xi )} \big )} = \sum_{i = 1}^l {\bE\big ( {g_i ( \xi )| {S_i} } \big )}
\bP({S_i} ).
$$

\section{Basic formula}

We consider the features vector $\xi( \omega)$: $\Omega\to H$ as a random
signal. Suppose $\mu$ is the probability distribution of the random signal
$\xi$. Let us define one linear form and two
bilinear forms for the random signal $\xi$
\beq
&&\langle{m,y} \rangle = \bE\langle{\xi ,y} \rangle = \int\limits_H
{\langle {x,y} \rangle \mu ( {\mathrm{d}x} )}, \nn \\
&&\langle{Ky,z} \rangle = \bE\Big ( {\langle {\xi ,y} \rangle \langle {\xi ,z}
\rangle} \Big ) = \int\limits_H {\langle {x,y} \rangle \langle {x,z} \rangle
\mu ( {\mathrm{d}x} )},   \\
&&\langle{Ry,z} \rangle = \bE\Big ( {\langle {\xi - m,y} \rangle \langle {\xi -
m,z} \rangle} \Big ) = \int\limits_H {\langle {x - m,y} \rangle \langle {x -
m,z} \rangle \mu ( {\mathrm{d}x} )}.
\eeq

A non-random signal $m$,  operator $K$,  and operator $R$ are called
a mathematical expectation, correlation operator, and covariance operator,
respectively.

    From (6) and (7), we have $\langle{Ky,z} \rangle =
\langle{Ry,z} \rangle + \langle{m,y} \rangle\langle{m,z} \rangle$. Then
$\langle{Ry,z} \rangle + \langle{m,y} \rangle\langle{m,z} \rangle =
\langle{( {R + p_m} )y,z} \rangle$, where $p_m y = \langle{y,m} \rangle
m$ is a one-rank operator.  It is evident that $p_m y = \| m \|^2 p_{\bar m} y$,  where $\bar m
= {m / {\| m \|}}$ and $p_{\bar m} y = \langle{y,\bar m} \rangle\bar m$
is a one-dimensional projection. Then
\beq
K = R + p_m = R + \| m \|^2 p_{\bar m}.
\eeq

Let the signal $x = \xi( \omega)$ be the pattern of the object $\omega$. An
affine structure of Hilbert space $H$ is used when realizations of
random signal is considered as points. Using a vector structure $H$,  it
is possible to interpret a value $\| x \|^2$   as a physical value, for
example, as energy, power, or intensity. The value $\| x \|^2$ is
a measure of deviation of signal from the zero vector, and nature  uses some
energy for this deviation. In the following, let this value be energy.

   Let $\langle{A\xi ,\xi} \rangle$ be a bilinear form, where $A$ is a linear
operator. Then
\beq
\bE\langle{A\xi ,\xi} \rangle = \int\limits_H {\langle {Ax,x} \rangle \mu
( {\mathrm{d}x} )} = \int\limits_H {\langle {x,Ax} \rangle \mu (
{\mathrm{d}x} )} = \mathrm{tr}KA = \mathrm{tr}AK.
\eeq

    If $P$ is an orthogonal projection, then $\langle{P{\kern 1pt} \xi ,\xi}
\rangle$ is the membership function. We can define a fuzzy event $A_P =
\{ {\omega , \langle {P{\kern 1pt} \xi ( \omega ),\xi ( \omega )}
\rangle \colon\ \omega \in \Omega} \}$.  From  (1) and (9), the probability of the fuzzy
event $A_P$ is defined as
$$
\bP( {A_P} ) = \bE\langle{P{\kern 1pt} \xi ,\xi}
\rangle = \int\limits_H {\langle {Px,x} \rangle \mu
(\mathrm{d}{\kern 1pt} x)} = \mathrm{tr}PK = \mathrm{tr}KP.
$$

   We now prove formula (9). Let $\{ {e_i} \}, i = 1 \ldots n,$ be an
orthonormal basis in $H$.  Using definitions of trace and correlation
operator (6), we have
\beqz
\mathrm{tr}KA &=& \sum_{i = 1}^n {\langle
{KAe_i ,e_i} \rangle} = \sum_{i = 1}^n {\int\limits_H
{\langle {x,Ae_i} \rangle \langle {x,e_i} \rangle
\mu ( {\mathrm{d}x} )}} = \int\limits_H {\sum_{i =
1}^n {\big \langle {A^* x,\langle {x,e_i} \rangle e_i}
\big \rangle} \mu ( {\mathrm{d}x} )} \\
&=&\int\limits_H
{\bigg \langle {A^* x,\sum_{i = 1}^n {\langle {x,e_i}
\rangle e_i}} \bigg \rangle} {\kern 1pt} \mu (
{\mathrm{d}{\kern 1pt} x} ) = \int\limits_H {\langle {A^*
x,x} \rangle} {\kern 1pt} \mu ( {\mathrm{d}{\kern 1pt} x}
) = \int\limits_H {\langle {x,Ax} \rangle {\kern 1pt}}
\mu ( {\mathrm{d}{\kern 1pt} x} ).
\eeqz
Since the scalar product is symmetric in a real Hilbert space,
$\langle{x,y} \rangle = \langle{y,x} \rangle$, we get $\langle{Ax,x}
\rangle = \langle{x,Ax} \rangle$. Then
\beqz
\mathrm{tr}AK &=& \sum_{i = 1}^n {\langle
{AKe_i ,e_i} \rangle} = \sum_{i = 1}^n {\langle {Ke_i
,A^* e_i} \rangle} = \sum_{i = 1}^n {\int\limits_H
{\langle {x,e_i} \rangle \langle {x,A^* e_i}
\rangle \mu ( {\mathrm{d}{\kern 1pt} x} )}} \\
&=&\int\limits_H {\bigg \langle {\sum_{i = 1}^n {\langle {x,e_i}
\rangle x} ,A^* e_i} \bigg \rangle \mu ( {\mathrm{d}{\kern
1pt} x} )} = \int\limits_H {\langle {x,Ax} \rangle \mu
( {\mathrm{d}x} )} = \int\limits_H {\langle {Ax,x}
\rangle \mu ( {\mathrm{d}x} )} = \bE\langle {A\xi
,\xi} \rangle.
\eeqz

Statistical states of quantum system are described by von Neumann
density matrix~[8]. In fact, von Neumann density matrix is the
correlation matrix of the discrete probability distribution. The formula (9)
enables to describe statistical states of quantum system with continuous
probability distributions.

\section{Recognition of two signal classes }

K. Helstrom  was first who considered  recognition of two classes    in the
quantum theory~[8]. We apply Helstrom's result for recognition of two
classes of random signals;  we only consider an energy distribution
instead of quantum probability distribution on projections.

    Assume that the object $\omega$ of recognition belongs to one of the
 classes $S_i$, $i = 1,2$, and the pattern of object is the signal
 $x = \xi(\omega)$. Suppose that each class $S_i$,  $i = 1,2$, is matched
 with the orthogonal projection $P_i$,  $i = 1,2$, where $P_1 + P_2 = I$.
 Then the value
 $\langle{P_i x,x} \rangle =
 \langle{P_i \xi ( \omega ),\xi ( \omega )} \rangle =
 g_i ( {\xi ( \omega )} )$ is the membership of
 object $\omega$ to the class $S_i$, $i = 1,2$. Therefore, the projections
 $P_i$, $i = 1,2$, define a class $\Phi$ of discriminant functions
 $g_i ( x ) = \langle{P_i x,x} \rangle $, $i = 1,{\kern 2pt} 2$.

     Let $p_i = \bP( {S_i} )$, $i = 1,2$ be a priori probabilities of classes
 and the conditional distributions $\mu_i ( A ) = \bP( {\xi \in A| {S_i}} )$,
 $i =1,2$, have the correlation operators $K_i$,  $i = 1,2$.
 We define a fuzzy event
 $G = \{ {\omega,  g( {\gamma ( \omega ),\xi ( \omega )} )
 \colon\ \omega \in \Omega} \}$, where
 $g( {\gamma, \xi} ) = \langle{P_\gamma \xi, \xi} \rangle$.
 By (4), we must maximize the probability of the fuzzy event $G$:
\begin{equation}
\label{eq10}
\bP( G ) = \bE g( {\gamma, \xi} ) = p_1 \int\limits_H
{\langle {P_1 x,x} \rangle} \mu_1 (\mathrm{d}x) + p_2
\int\limits_H {\langle {P_2 x,x} \rangle} \mu_2 (\mathrm{d}x).
\end{equation}

    Let us suggest an energy interpretation of formula (10). Using (5) and
 (10), we have
$$
r_j ( i ) = \bE\big ( {1_{S_j} \langle {P_i \xi, \xi} \rangle} \big ) = p_j
\int\limits_H {\langle {P_i x,x} \rangle} \mu_j ( {\mathrm{d}{\kern 1pt}
x} ) = p_j \mathrm{tr}P_i K_j.
$$
 Each projection $P_i, {\kern 1pt} i = 1,{\kern 1pt} 2{\kern 1pt}$,  passes
 same energy of signals $x = \xi( \omega)$ from the own class
 $S_j, {\kern 1pt} i = j$ and the other class $S_j, {\kern 1pt} i \ne j$.
 We call energy $r_j ( i )$ a ``correct'' energy if $i = j$ and an ``error'' energy if $i \ne j$. We also call a full ``correct'' energy, which passes
 projections of all classes, as an energy of ``correct'' recognition.
 This energy is defined as
\beq
\mathrm{Enr}_\mathrm{C} ( {P_1,  P_2} ) = r_1 ( 1 ) + r_2 ( 2 ) = p_1
\mathrm{tr}P_1 K_1 + p_2 \mathrm{tr}P_2 K_2.
\eeq
 It is clear that we must find projections $P_1,  P_2$ so that the value
 $\mathrm{Enr}_\mathrm{C} ( {P_1,  P_2} )$ would be the largest. In
 other words, projections $P_1,  P_2$ together must pass the energy of
 signals from their own classes as much as possible.

    Since  $P_2 = I - P_1$,  we have
\[
\mathrm{Enr}_\mathrm{C} ( {P_1,  P_2} ) = p_2 \mathrm{tr}K_2
+ \mathrm{tr}P_1 ( {p_1 K_1 - p_2 K_2} ).
\]
Here the first value is constant but the second value depends only on
the projection $P_1$.  Hence we must find the projection $P_1$ such that
the second value was the largest. Assume that $\lambda_i,  i =
1\ldots n$, are eigenvalues and $y_i,  {\kern 1pt} i = 1\ldots n$, are the
eigenvectors of the operator $p_1 K_1 - p_2 K_2$.  Then
\beqz
\mathrm{tr}P_1 ( {p_1 K_1 - p_2 K_2} ) &=&
\sum_{i = 1}^n {\big \langle {P_1 ( {p_1 {\kern 1pt} K_1
- p_2 K_2} )y_i, y_i} \big \rangle} = \sum_{i = 1}^n
{\langle {P_1 {\kern 1pt} \lambda _i y_i, y_i} \rangle}\\
 &=&\sum_{i = 1}^n {\lambda _i} ||P_1 y_i ||^2 = \sum_{\lambda
_i {\kern 1pt} > 0} {\lambda _i ||P_1 y_i ||^2} + \sum_{\lambda
_i \le 0} {\lambda _i ||P_1 y_i ||^2} = d_1 + d_2,
\eeqz
where $\| {P_1 y_i} \|^2 \le\| {y_i} \|^2$ for all $i = 1\ldots n$,  $d_1 > 0$,  $d_2 \le 0$.  Let $P_1$ be a projection onto a subspace
spanned by the eigenvectors with positive eigenvalues. Then $\| {P_1 y_i}
\|^2 = \| {y_i} \|^2$ if $\lambda_i > 0$ and $\| {P_1 y_i} \|^2 = 0$ if
$\lambda_i \le 0$.  It follows that $d_1$ will be the largest  and $d_2
= 0$.  Hence the required projection $P_1$ is found and $P_2 = I - P_1$.

\smallskip
\noindent \textsc{Comment 1.} It is possible to minimize the energy of ``error''
recognition. The energy of ``error'' recognition is the following sum:
$$
\mathrm{Enr}_\mathrm{E} ( {P_1,  P_2} ) = p_1 r_1 ( 2 ) + p_2 r_2 ( 1 )
= p_1 \mathrm{tr}P_2 K_1 + p_2 \mathrm{tr}P_1 K_2.
$$

   If the projections $P_1,  P_2$ maximize the energy of ``correct''
recognition, then they must minimize energy of  ``error''
recognition. Indeed, we have
\beq
\mathrm{Enr}_\mathrm{E} ( {P_1,  P_2} ) &=& p_1
\mathrm{tr}(P_2 K_1 ) + p_2 \mathrm{tr}(P_1 K_2 )= p_1 \mathrm{tr}(I -
P_1 )K_1 + p_2 \mathrm{tr}( {I - P_2} )K_2  \nn \\
&=& p_1\mathrm{tr}K_1 + p_2 \mathrm{tr}(K_2 ) - p_1 \mathrm{tr}P_1 K_1 - p_2\mathrm{tr}(P_2 K_2 ) \nn \\
\quad {}&=& p_1 \mathrm{tr}K_1 + p_2 \mathrm{tr}(K_2 ) -
\mathrm{Enr}_\mathrm{C} ( {P_1,  P_2} ).
\label{eq12}
\eeq
There the values $p_1 \mathrm{tr}K_1$ and $p_2 \mathrm{tr}K_2$ are
constant. Hence the value $\mathrm{Enr}_\mathrm{E} ( {P_1,  P_2} )$
will be the least if the value $\mathrm{Enr}_\mathrm{R} ( {P_1,  P_2} )$
is the greatest.

\smallskip
\noindent \textsc{Comment 2.} From (12) it follows that the sum energy of ``correct''
recognition and ``error'' recognition is a constant. Thus, increasing
the energy of ``correct'' recognition, we decrease the energy of
``error'' recognition and vice versa.

\section{Decision rule for recognition}

Suppose there are two classes of objects $S_i,  i = 1,2$, and the signal
$x = \xi( \omega)$ is the pattern of the object $\omega$. If we use a
probabilistic Bayesian classifier, then the feature space $H$ is divided
into the disjoint subsets: $L_1$, $L_2$,  $L_1 \cup L_2 =
\mathrm{H}$,  where the subset $L_1$ correspond to the class
$S_1$ and the subset $L_2$ corresponds to the class $S_2$.  The decision
rule that determines unambiguously  to which class $S_1$ or $S_2$
belongs the object $\omega$, is defined as follows: $\omega\in S_1$ if
$x \in L_1$ and $\omega\in S_2$ if $x \in L_2$.

     However, the situation is different when quantum logic is used. Suppose
 each class $S_i$,  $i = 1,2$, is matched with the orthogonal projection
 $P_i$, $i = 1,2$, where $P_1 + P_2 = I$. Denote $L_1 = P_1 H$, $L_2 = P_2 H$,
 where $L_1 \oplus L_2 = H$. Then the pattern of the object
 $x = \xi( \omega)$ can be a sum of two signals: $x = P_1 x + P_2 x = x_1 + x_2$,
 where $x_1 \in L_1$,  $x_2 \in L_2$. It is natural to accept that
 $\omega\in S_1$ if $P_1 x = x$ and $\omega\in S_2$ if $P_2 x = x$. If
 $x_1 \ne 0$ and $x_2 \ne 0$, then the pattern $x$ belongs
 simultaneously to two subspaces: $L_1$ and $L_2$. Hence we can not
 decide to which class belongs the object using subspaces of quantum logic. Therefore we must use discriminant functions
 $g_i ( x ) = \langle{P_i x,x} \rangle, i = 1,2$, which unambiguously
 gives the decision about the membership of the object to one of
 the classes: $S_1$ or $S_2$. By (11), we can find discriminant functions
 $g_1 ( x ) = \langle{P_1 x,x} \rangle$ and $g_2 ( x ) =
 \langle{P_2 x,x} \rangle$
 such that they maximize the energy of ``correct'' recognition.
 Thus we have the following decision rule:
\beq
\omega\in S_1\ \ \mbox{if}\ \ \langle{P_1 x,x} \rangle > \langle{P_2 x,x}
\rangle \ \ \mbox{and}\ \ \omega\in S_2\ \ \mbox{otherwise.}
\eeq

      When the decision rule (13) is applied, the feature space $H$ is
 divided into disjoint sets:
 $A_1 = \{ {x\colon\ \langle {P_1 x,x} \rangle >
 \langle {P_2 x,x} \rangle} \}$ and $A_2 = \{ {x\colon\ \langle {P_2 x,x}
 \rangle \ge \langle {P_1 x,x} \rangle} \}$. We put
$$
\mathrm{Enr}_\mathrm{C} ( {A_1, A_2} ) = p_1
\int\limits_{A_1} {\langle {P_1 x,x} \rangle} \mu_1 (
{\mathrm{d}{\kern 1pt} x} ) + p_2 \int\limits_{A_2} {\langle
{P_2 x,x} \rangle} \mu_2 ( {\mathrm{d}{\kern 1pt} x} ).
$$
It is evident that
\beq
\mathrm{Enr}_\mathrm{C} ( {P_1, P_2} ) = \mathrm{Enr}_\mathrm{C} ( {A_1
,A_2} ) + p_1 \int\limits_{A_2} {\langle {P_1 x,x} \rangle} \mu_1 (
{\mathrm{d}x} ) + p_2 \int\limits_{A_1} {\langle {P_2 x,x} \rangle}
\mu_2 ( {\mathrm{d}x} ).
\eeq

The object $\omega$ of recognition is chosen in a random way but we hope that
the value of the discriminant function $g_i ( x )$ of class $S_i$ is maximal
if statement $\omega\in S_i$ is true.  Also it is natural to hope that
$\mathrm{Enr}_\mathrm{C} ( {P_1, P_2} )$ is approximately equal  to
$\mathrm{Enr}_\mathrm{C} ( {A_1, A_2} )$. Using $\langle{P_1 x,x}
\rangle\le\langle{P_2 x,x} \rangle$ on $G_2$ and $\langle{P_2 x,x}
\rangle < \langle{P_1 x,x} \rangle$ on $G_1$,  we get
\beqz
&&p_1 \int\limits_{A_2} {\langle {P_1 x,x}
\rangle} \mathrm{\mu  }_1 ( {\mathrm{d}x} ) \le p_1
\int\limits_{A_2} {\langle {P_2 x,x} \rangle} \mathrm{\mu
}_1 ( {\mathrm{d}x} ) \le p_1 \int\limits_H {\langle
{P_2 x,x} \rangle} \mathrm{\mu  }_1 ( {\mathrm{d}x} ) =
p_1 \mathrm{tr}P_2 K_1,  \\
&&p_2 \int\limits_{A_1} {\langle {P_2
x,x} \rangle} \mathrm{\mu  }_2 ( {\mathrm{d}{\kern 1pt} x}
) \le p_2 \int\limits_{A_1} {\langle {P_1 x,x} \rangle}
\mathrm{\mu  }_2 ( {\mathrm{d}{\kern 1pt} x} ) \le p_2
\int\limits_H {\langle {P_1 x,x} \rangle} \mathrm{\mu  }_2
( {\mathrm{d}{\kern 1pt} x} ) = p_2 \mathrm{tr}P_1 K_2.
\eeqz
From (14) it follows that
\beq
\label{eq1 5}
0 \le\mathrm{Enr}_\mathrm{C} ( {P_1, P_2} ) -
\mathrm{Enr}_\mathrm{C} ( {A_1, A_2} ) \le p_1 \mathrm{tr}P_2
K_1 + p_2 \mathrm{tr}P_1 K_2 = \mathrm{Enr}_\mathrm{E} ( {P_1, P_2}
).
\eeq
If projections $P_1,  P_2$ maximize the energy $\mathrm{Enr}_\mathrm{C} (
{P_1, P_2} )$ of ``correct'' recognition, then from comment 1 it follows
that projections $P_1,  P_2$ minimize the energy $\mathrm{Enr}_\mathrm{E}
( {P_1, P_2} )$ of ``error'' recognition. If we have good recognition
with projections $P_1,  P_2$,  then the value $\mathrm{Enr}_\mathrm{E} (
{P_1, P_2} )$ is small. Therefore from (15) it follows that
$\mathrm{Enr}_\mathrm{C} ( {P_1, P_2} )$ is  approximately equal to
$\mathrm{Enr}_\mathrm{C} ( {A_1, A_2} )$.

\medskip
\noindent \textsc{Example 1.}
Suppose the object of recognition $\mathrm{\omega  }$ belongs
to one of the classes $S_i$,  $i = 1,2$.  Assume that a priori probabilities
of classes are equal $p_{1} = p_{2} = {1 / 2} ;$  the conditional
distributions $\mu_i ( A ) =\bP( {\xi \in A| {S_i}} )$, $i = 1,2$,  have the
identical covariance matrices equal to $R$ and mathematical expectations
$m_{1}$,  $m_{2}$ are orthogonal as vectors.

We choose the orthonormal basis $e_i$,  $i = 1\ldots n$, in $H$
such that $e_{1} = {m_1 } / {\| {m_1} \|}$, $e_{n} = {m_2 } / {\| {m_2}
\|}$.   We get  from (8) that $K_1 = R + \| {m_1} \|^2 p_1$,
$K_2 = R + \|{m_2} \|^2 p_2$,  where $p_1 x = \langle{x,e_1} \rangle e_1$,
$p_2 x =\langle{x,e_n} \rangle e_n$.  In the chosen basis, the matrix $p_1 K_1 -
p_2 K_2 = {1 / 2}( {K_1 - K_2} )$ is diagonal with eigenvalues ${{\|
{m_1} \|^2} / 2},0,\ldots, 0, - {{\| {m_2} \|^2} / 2}$.  Then $P_1 x =
{\langle {x,m_1} \rangle } / {\| {m_1} \|}$,  $P_2 x = {\langle {x,m_2}
\rangle } / {\| {m_2} \|}$.  If $x = \xi( \omega)$ is the pattern of
the object $\omega$, then by (13) we have the following decision rule:
$\omega\in S_1$ if ${\langle {m_1, x} \rangle ^2 } / {\| {m_1} \|^2} >
{\langle {m_2, x} \rangle ^2 } / {\| {m_2} \|^2}$ and $\omega\in S_2$
otherwise.

\section{Normalization by trace}

Suppose $x = \xi( \omega)$ is the pattern of the object $\omega$ and $\bE( {
{\langle {P\xi, \xi} \rangle} |S_i} ) = \mathrm{tr}PK_i$,  $i = 1,2$, are
 conditional energy distributions on projections. The conditional energy
distributions on projections of different classes are not equivalent if
the trace of the correlation operators $K_i$,  $i = 1,2$, are not equal. It is
possible to normalize the conditional energy distribution on projections
by normalizing the pattern of objects of each class as follows: $\eta_i =
{\xi / {\sqrt{\mathrm{tr}K_i},\  i = 1,2}}$.  Then the correlation
operators will be normalized as follows: $\bar K_1 { = K_1 } /
{\mathrm{tr}K_1}$, $ \bar K_2 { = K_2 } / {\mathrm{tr}K_2}$,  where
$\mathrm{tr}\bar K_1 = \mathrm{tr}\bar K_2 = 1$.  Also it is necessary
to normalize the object patterns $x = \xi( \omega)$ in the decision
rule (13). So, we have the following decision rule: $\omega\in S_1$ if ${\langle
{P_1 x,x} \rangle } / {\mathrm{tr}K_1} > {\langle {P_2 x,x} \rangle } /
{\mathrm{tr}K_2}$ and $\omega\in S_2$ otherwise.

\medskip
\noindent \textsc{Example 2.}
 We consider a classical recognition task of two classes: the
class $S_1$ is a random signal $\xi = a + \eta$, where $a$ is a non-random
signal and $\eta$ is a white noise; the class $S_2$ is a white noise
$\eta$. Suppose $p_{1} = p_{2} = {1 / 2}$.

The correlation matrix of white noise $\eta$ is $\sigma^2 I$,  where
$\sigma^2$ is a constant and $I$ is an identity matrix. The mathematical
expectations of the random signals of classes $S_i,  i = 1,2$, are
respectively $m_{1} = a$, $m_{2} = 0$.  Applying the decision rule of
example 1, the classifier always decide that all objects $\omega\in
S_1$.

    We normalize the correlation matric\textsl{}es of both classes by their trace.
From (8), we have $K_1 = \sigma^2 I + \| a \|^2 p_{\bar a}$,  where
$p_{\bar a} x = \langle{x,\bar a} \rangle\bar a$,  $ \bar a = {a / {\| a
\|}} ;$ we also  have $K_2 = \sigma^2 I$.  Then $\mathrm{tr}K_1 =
\sigma^2 \mathrm{tr}I + \| a \|^2 \mathrm{tr}p_{\bar a} = n\sigma^2 + \|
a \|^2$ and $\mathrm{tr}K_2 = n\sigma^2$. Since  covariance matrices of both
classes are $\sigma^2 I$,
 they are diagonal in any basis. We choose the basis in $H$ such that
 $e_{1} = \bar a$. Then the matrix
 ${p_1 \bar K_1 - p_2 \bar K_2}  = {{1 / 2}( {{K_1 } / {\mathrm{tr}}K_1 -
 {K_2 } / {\mathrm{tr}}K_2} )}$
 is diagonal in the chosen basis with following eigenvalues:
$$
\frac{( {n - 1} )\| a \|^2} {2n( {n\sigma ^2 + \| a \|^2} )}, \cdots, -
\frac{\| a \|^2} {2n( {n\sigma ^2 + \| a \|^2} )}, - \frac{\| a \|^2}
{2n( {n\sigma ^2 + \| a \|^2} )}.
$$
 Here the first eigenvalue is positive and the last $n - 1$ eigenvalues are
 negative. So the projection $P_1$ is a one-dimensional projection:
 $P_1 x = \langle{x,e_{1}} \rangle e_{1}$. Then
 $\langle{P_1 x,x} \rangle = {\langle {x,a} \rangle ^2 } / {\| a \|}^2$ and
 $\langle{P_2 x,x} \rangle = \langle{( {I - P_1} )x,x} \rangle =
 \langle{x,x} \rangle - {\langle {x,a} \rangle ^2 } / {\| a \|}^2$.
 By (9), the variance of the white noise $\eta$ is equal to
 $\bE\| \eta\|^2 = \bE\langle{\eta, \eta} \rangle = n\sigma^2$.
 So the signal-to-noise ratio is defined as
 ${\mathrm{SNR} = \| a \|^2} / {n\sigma ^2}$.

    Normalizing the object pattern by the trace, we
 get from (13) the following  decision rule: $\omega\in S_1$ if
 ${\langle {x,a} \rangle ^2 } / {( {1 + \mathrm{SNR}} )} >
 \| x \|^2 \| a \|^2 - \langle{x,a} \rangle^2$ and
 $\omega\in S_2$ otherwise.

    We have $\mathrm{tr}P_2 K_1 = \sigma^2 \mathrm{tr}P_2 = ( {n - 1})\sigma^2$ and
 $\mathrm{tr}P_1 K_2 = \sigma^2 \mathrm{tr}P_1 = \sigma^2$.
 Then
$$
\mathrm{Enr}_\mathrm{E} ( {P_1, P_2} ) = \frac{p_1} {\mathrm{tr}K_1}
\mathrm{tr}P_2 K_1 + \frac{p_2} {\mathrm{tr}K_2} \mathrm{tr}P_1 K_2 =
\frac{( {n - 1} )\sigma ^2} {2( {n\sigma ^2 + \| a \|^2} )} +
\frac{\sigma ^2} {2n\sigma ^2} = \frac{1 - {1}/{n}}{2( {1 +
\mathrm{SNR}} )} + \frac{1}{2n}.
$$
 Thus the energy of ``error'' recognition is small if the $\mathrm{SNR}$ and
 the dimension $n$ of the feature space $H$ are large.

\section{Normalization by signal norm}

    We can to normalize object pattern by normalizing each signal
 $x = \xi( \omega)$ as vector by its norm. In that case, ends of
 normalized random vectors are located on a unit sphere. Suppose
 $ \bP( {\xi = O} ) = 0$. Putting $\eta = {\xi / {\| \xi \|}}$, we have
\beq
\bE\langle{\eta, \eta} \rangle \!=\! \bE\big ( {{\langle {\xi, \xi} \rangle} / {\|\xi \|^2}} \big )\!=\! \bE\big ( {{\|\xi \|^2} / {\|
\xi \|^2}} \big )\! = 1.\quad
\eeq

    Let $\bar K$ be the correlation operator of the normalized random
 signal $\eta$. From (9) and (16), we have $\mathrm{tr}\bar K = 1$.
 Hence, the energy distribution on projections is normalized.

    If objects patterns of are normalized as
 $x = {\xi ( \omega )} / {\| {\xi ( \omega )} \|}$, then
 $g_i ( x ) = \langle{P_i x,x} \rangle\le 1$, $i = 1,2$.
 This yields that $\sup g_i ( x ) = 1$, where $i = 1,2$.
 So the discriminant functions $g_i ( x ), i = 1,2$
 are classical membership functions [1].

    Vectors $x$ and $\lambda x$ for any $\lambda\mathrm{ > 0}$ describe
 the same physical state in quantum mechanics. It means that states of
 quantum systems are rays, i.e. points of projective space. Due this
 fact, we can consider  states with unit norm $\| x \| = 1$ only.

    The same holds for sound signals and monochrome images. In fact, the
 sound signals $x$ and $\lambda x$ for any $\lambda\mathrm{ > 0}$ differ
 in loudness only. The monochrome images can be described as a set of $l = nm$
 real numbers corresponding to the intensity of the light in each pixel.
 Hence the space of the monochrome images can be described as a vector
 space of dimension $l = nm$. All the intensities of the monochrome image
 can be multiplied by a number $\lambda\mathrm{ > 0,}$
 but that does not change monochrome image.

\section{Subtraction of mean}

    The following hypothesis is accepted in the recognition theory:
 the distribution of the patterns of a class is concentrated in a compact
 area of feature space. It is natural to assume that distribution of patterns
 is grouped around the mean (mathematical expectation) of this distribution.
 Then each object pattern  $x = \xi(\omega)$ can be written as the sum
 $x = y + a$, where $a$ is the mean and $y$ is  the random vector from
 the compact area such that its beginning is the end of the mean $a$.

 On the other hand, linear subspaces that correspond to classes in feature space are intersect at the zero point of the space ~$H$ (the origin of the coordinates). Therefore if quantum logic is used for recognition, then it is natural to combine compact areas with the origin of coordinates.

   In this case, the energy distributions on projections are described by
 the covariance operators.

    Suppose the conditional distributions
 $\mu_i ( A ) = \bP( {\xi \in A|{S_i}})$, $i = 1,2$,
 have the covariance operators $R_1,  R_2$ and means
 $m_1,  m_2$. Then it is necessary to find projections $P_1,  P_2$
 such that the value of energy
 $\mathrm{Enr}_\mathrm{R} ( {P_1,  P_2} ) = p_1 \mathrm{tr}P_1 R_1 + p_2 \mathrm{tr}P_2 R_2$
 would be the maximal. After subtracting from object patterns
 $x = \xi( \omega)$ their means, we get from (13) the following decision rule:
 $\omega\in S_1$
 if
 $\langle{P_1 ( {x - m_1} ),x - m_1} \rangle > \langle{P_2 ( {x - m_2} ),x - m_2} \rangle$
 and
 $\omega\in S_2$ otherwise.

\end{document}